\documentclass[aps,pra,twocolumn,nopacs,superscriptaddress,floatfix]{revtex4}

\usepackage{amsmath} \usepackage{bbm}
\usepackage{epsfig}

\newcommand{\id}{\mathbbm{1}}

\usepackage{times}

\newcommand{\cc}{{\mathbbm{C}}}
\newcommand{\rr}{{\mathbbm{R}}}
\newcommand{\nn}{{\mathbbm{N}}}
\newcommand{\me}{\mathrm{e}}
\newcommand{\mi}{\mathrm{i}}

\newtheorem{theorem}{Theorem}

\newtheorem{lemma}[theorem]{Lemma}

\begin{document}

\title{Supersonic quantum communication}

\author{J.\ Eisert}

\affiliation{Institute of Physics and Astronomy, University of Potsdam, 14476 Potsdam,
Germany}
\affiliation{Quantum Optics and Laser Science, 
Imperial College London,
London SW7 2PE, UK}

\author{D.\ Gross}
\affiliation{Quantum Optics and Laser Science, 
Imperial College London,
London SW7 2PE, UK}

\affiliation{
Institut f\"ur Mathematische Physik, Technische Universit\"at
Braunschweig, 
38106 Braunschweig, Germany
}

\begin{abstract}
	When locally exciting a quantum lattice model, the excitation will
	propagate through the lattice.  The effect is responsible for a
	wealth of non-equilibrium phenomena, and has been exploited to
	transmit quantum information through spin chains.  
	It is a commonly
	expressed belief that for local Hamiltonians, any 
	such propagation
	happens at a finite ``speed of sound''. Indeed, 
	the Lieb-Robinson
	theorem states that in spin models, all effects 
	caused by a perturbation
	are limited to a causal cone defined by a constant speed, up to
	exponentially small corrections.  In this work we show that	
	for translationally invariant bosonic models with
	nearest-neighbor interactions, this belief is incorrect: We prove
	that one can encounter excitations which accelerate under the
	natural dynamics of the lattice and allow for reliable transmission
	of information faster than any finite speed of sound. The effect is
	only limited by the model's range of validity (eventually by
	relativity). It also implies that in non-equilibrium dynamics of
	strongly correlated bosonic models far-away regions may 
	become quickly entangled, suggesting that their simulation
	may be much harder than that of spin chains 
	even in the low energy sector.
\end{abstract}	

\maketitle

Quantum spin chains---or more generally, quantum spin models on a
lattice---are ubiquitous in condensed matter physics and quantum
optics. They share the fundamental feature that perturbations will
propagate through the lattice at some characteristic ``speed of
sound''  \cite{LR,Travel}. This effect plays an
important role for a wealth of non-equilibrium phenomena in many-body
systems, e.g., for the dynamics of relaxation processes towards
equilibrium \cite{Quench,Quench2}.  In the context of quantum
information science, it has been noted that excitations propagating
through a spin chain may be used to transmit quantum
information---thus turning a spin chain into a quantum channel. Here,
the appealing feature is that the transport is not facilitated by
engineered quantum gates, but rather by the natural time evolution of
the lattice system \cite{Bose,Harmonic,Christandl}.

Because in lattice models only neighboring systems interact with each
other directly, it is intuitive to assume that the maximal propagation
speed of excitations (i.e., the speed of sound) is finite and given by
a value characteristic for each model. Indeed, an analogous statement
is clearly true for relativistic systems, where a perturbation can
have no influence outside its causal cone. Mathematical physics
provides a rigorous justification for this observation in the form of
Lieb-Robinson bounds \cite{LR}: in spin lattice systems, perturbations
can spread only linearly in time, up to exponentially small
corrections. Recently, an analogous result has been proven to
hold for a class of bosonic systems \cite{Schlein}. 

Interestingly, familiar as the belief that propagation of excitations
in local models happens with a finite velocity may be: it is not quite
right. We demonstrate that certain well-defined local bosonic models
allow excitations to accelerate to arbitrarily high velocities.  The
effect is only limited by the range of validity of the model (which
must certainly break down with the onset of relativistic effects).  It
occurs even for single excitations with bounded energy, traveling
along a one-dimensional chain of bosons with translationally invariant
nearest-neighbor interactions.  From the quantum information
perspective, we show that the quantum channel associated with this
chain has a strictly positive information capacity, even after a time
sub-linear in the length of the chain.   While the presented models
are non-integrable, we derive the results rigorously, without
resorting to numerical means. We do so by considering single
excitation spaces and---in this context unusual---invoke ideas from
convex optimization.

\begin{figure}
\includegraphics[width=5.3cm]{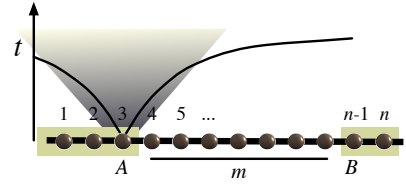}
\caption{\label{sus}
In quantum spin chains, local excitation will travel no faster than
some characteristic speed of sound. The phenomenon defines a causal
cone, outside of which any influence is exponentially
suppressed. For bosonic models, it is demonstrated that the causal
region may be bent to the outside, covering distances exponential in
time. Information can propagate from region $A$ to $B$ at arbitrary
velocities, via the quantum channel defined by the free time
evolution of the chain. Note that this contrasts the situation in
disordered spin chains, where the causal region is curved to the inside
\cite{Tobias}.}
\end{figure}

There are several conceptual consequences of these results. It is now
clear that any analysis of non-equilibrium processes in bosonic models
must incorporate the possibility of far-away regions exchanging
information on short time scales. In particular, it seems likely that
simulating short-term dynamics even of the low-energy sector of
bosonic models is much harder than for spin chains (where
Lieb-Robinson bounds are the basis for efficient algorithms
\cite{Travel,TobiasPaper}). Further, the results highlight the
non-triviality of Lieb-Robinson bounds for spin chains with
finite-dimensional constituents.

More practically---while the models we present very strongly violate
any bound on propagation speeds---they have reasonable physical
properties. Related models with similar features could well be
realized by tuning the parameters of suitable physical systems, for
example in arrays of coupled cavities with polariton excitations \cite{Cavities}. 
This opens up the possibility of observing
accelerating excitations experimentally and, potentially, of using
bosonic chains as fast channels for quantum communication.

{\it Local Hamiltonians and causality in spin chains. --} 
A {\it local Hamiltonian} on $n$ sites is of the form
\begin{equation}\label{hamiltonian}
	H= \sum_{j=1}^n  h_j,
\end{equation}
where $h_j$ acts non-trivially only on a finite number of adjacent
sites. In what follows, we will restrict attention to the most
relevant case of nearest-neighbor interactions.  Quantum information
transmission through spin chains with Hamiltonians as above
has been extensively studied in the literature. 

Before turning to bosonic models, let us first recall the precise
situation for spin chains ($d$-level systems). The fact that there
always exists a speed of sound---a maximal speed of information
propagation---is the content of the following {\it Lieb-Robinson bound}
\cite{LR}: If $A$ and $B$ are operators which
act non-trivially only on some (distinct) regions of the chain, then
there exist constants $\mu, C>0$ and a velocity $v>0$ such that
\begin{equation}\label{so} 
	\|[ A(t), B]\|\leq C \| A\|\, \| B\|\,
	e^{-\mu (\operatorname{dist}(A,B)-v|t|)}, 
\end{equation} 
for all times $t$.  Here, $ A(t)= e^{\mi Ht}  A e^{- \mi  Ht}$ is the
time-evolved observable, $\|.\|$ the operator norm, and
$\operatorname{dist}(A,B)$ denotes the number of sites between the
supports of $A$ and $B$ (see Fig.~\ref{sus}). The above form may seem
somewhat awkward at first sight. To get a more physical statement, one
may verify that Eq.~(\ref{so}) implies that any effect a perturbation
$A$ can have on a distant observable $B$ is exponentially suppressed
outside the causal cone defined by
$|t|\geq\operatorname{dist}(A,B)/v$.  In particular, any
non-exponentially suppressed quantum communication using this spin
chain can happen at most with velocity $v$ \cite{Travel}.
Due to the intuitive nature of this statement the above bound is
often taken for granted or even dismissed as being ``trivial''.

{\it Supersonic communication. --} Roughly, we 
say that a model allows
for ``supersonic'' communication, if its dynamics can carry
information over distances $m$ in time $t(m)$ which scales
sub-linearly in $m$.  We will make this concept precise below.

The setting is a chain of $n$ bosonic systems with nearest-neighbor
interactions and open boundary conditions. The interactions should be
translationally invariant ($h_i=h_j$ in Eq.\ 
(\ref{hamiltonian})), up to 
the obvious modifications at the boundary.
We refer to the left sites
$1,\dots, a$ as section $A$ of the chain, whereas
sites $a+m,\dots, n$ form part $B$.
We assume that the system is initially in some
factoring, translationally invariant pure state 
$|\psi\rangle=|\psi_0\rangle^{\otimes n}$.
A party in control of region $A$ may now try to
communicate with a party at $B$ by either creating some excitations in
her end of the chain, or else leaving the system untouched. More
precisely, in the first case party $A$ would apply a unitary operator $U_A$ to region $A$. At the receiving end, party
$B$ waits for some time $t$ before probing whether a signal
corresponding to some POVM element $O_B$ is detected.
The statistics are influenced by $A$'s decision and given by
\begin{equation*}
	P_1= \text{tr}[O_B e^{-\mi 
	tH}U_A|\psi\rangle\langle\psi|U_A^\dagger e^{\mi
	tH}]
\end{equation*}
in case $A$ has excited the chain and
\begin{equation*}
	P_0=\text{tr}[O_B e^{-\mi
	tH}|\psi\rangle\langle\psi| e^{\mi
	tH}]
\end{equation*}
in case $A$ has not done so. The classical information capacity of the
channel thus defined is a function of the \emph{signal strength}
$\delta=|P_0-P_1|$. If $\delta$ scales as $1/\operatorname{poly}(m)$,
standard protocols involving polynomially many channels used in
parallel may be employed to, say, ``transmit radio signals through the
quantum chain'' with arbitrarily high fidelity.
The last relevant quantity is the energy scale of the states involved,
measured, e.g., by the variances
$E_0^2 = {\langle\psi|H^2|\psi\rangle}$,
$E_1^2 = {\langle\psi|U_A^\dagger H^2 U_A|\psi\rangle}$.
Low values for $E_0, E_1$ imply that the states are largely contained
in the low energy sector of $H$ \cite{ChebRemark}. 
Set $\varepsilon=\operatorname{max}\{E_0,E_1\}$.
Below, we define three increasingly 
strong ways in which bosonic models could potentially 
violate finite bounds on the maximum
propagation speed for signals. We go on to establish the main result: 
even the strongest scenario can be realized by
reasonable Hamiltonians.

(i) {\it Models which allow for arbitrarily fast transmission of
information, using polynomial resources}. More precisely, for every
signal velocity $m/t$, there should be suitable encoding operations
$U_A(m)$ and observables $O_B(m)$ such that the signal strength
$\delta(m)$ is of order $1/\operatorname{poly}(m)$. To obtain a
reasonable protocol, the energy scale $\varepsilon(m)$ 
should grow only polynomially in $m$. 
While models of this type are
interesting objects of study, it may be argued that their existence
would  not be too surprising. Indeed, as energy and time take
reciprocal roles in quantum mechanics, it is plausible that adding
``more energy'' to the system may lead to faster dynamics.  This
motivates the next, more stringent, situation.

(ii) {\it Models for which the signal velocity scales faster than the
inverse energy}. In addition to the definitions above, we demand that 
	${m\delta(m)}/{(t\varepsilon(m))}\to\infty$ 
as $m\to\infty$. For such
models, the phenomenon cannot just be explained by the fact that
unbounded Hamiltonians allow for signals with higher energies and thus faster dynamics.

In scenario (i), (ii) above, information propagates at arbitrarily
high velocities---yet the distance covered is still linear in time
(so the causal regions are cones with arbitrarily wide opening angles).
The final situation is more demanding, requiring that excitations
``speed up'' as they propagate. 

(iii) {\it Models allowing for accelerating signals}. Here, we require
that the signal strength $\delta$, the energy scale $\varepsilon$ and, in
fact, the encoding operation $U_A$ do not depend on the 
distance $m$, while the time $t$ should scale sub-linearly in $m$.
In the next section, we discuss situations exhibiting behavior of type
(iii) (and hence also of types (i,ii)) in a quite radical fashion.

{\it Models. --} 
The type of models we subsequently allow
for are governed by nearest-neighbor Hamiltonians of the form
\begin{equation*}
	H=
	\sum_{j=1}^{n-1} f_{j,j+1} + \sum_{j=1}^n g_j
\end{equation*}
with interaction term $f_{j,j+1}$ and on-site term $g_j$ .  For
$f_{j,j+1}=(a_j^\dagger +a_j)(a_{j+1}^\dagger+a_{j+1})$, 
and $g_j=\mu a_j^\dagger a_j$ for $\mu>0$ this is an
instance of a {\it harmonic chain}.   
For an
on-site interaction $g_j=\mu a_j^\dagger a_j + U
a_j^\dagger a_j (a_j^\dagger a_j-1)$ and a hopping
$f_{j,j+1}=J (a_j^\dagger   a_{j+1}+h.c.)$
this gives rise to  the {\it
Bose-Hubbard model}. 

Let us spend some time to develop the physical intuition behind the
constructions below. Consider a Hamiltonian with harmonic hopping
terms of the form $f_{j,j+1}=J (a_j^\dagger a_{j+1}+h.c.)$ and initial
state $|\psi\rangle=|0,\dots,0\rangle$  (using
the Fock basis).  Diagonalizing the operators $f_{j,j+1}$, we see that
the coupling strength between the sites grows as higher Fock layers
become populated. Now consider an on-site interaction $g_j$ which does
not preserve the Fock basis. If $g_j$ is e.g.\ a (low-order)
polynomial in $a_j, a_j^\dagger$, then any application of $g_j$ will
couple the state $|\psi(t)\rangle$ to higher and higher Fock layers.
For the sake of the argument, one may be tempted to model the spectrum
of local reductions  to undergo a random walk on the local
Fock basis, starting at $|0\rangle$ and gradually spreading to higher
levels. Therefore, for such Hamiltonians, the coupling strength
between sites could be expected to grow as time proceeds.  
Terms coupling different Fock states are common in physical models (as
long as the Fock basis does not represent massive particles). An
obvious example is provided by the harmonic chain 
$f_{j,j+1}=(a_j^\dagger +a_j)(a_{j+1}^\dagger+a_{j+1})$, whose
Hamiltonian contains terms of the form $a_j a_{j+1}$.
In particular, as Hamiltonian dynamics
by definition conserves energy, these terms do not inject energy
into the system.

It is unclear at this point whether this increase in coupling strength
gives rise to an acceleration of the dynamics. Our objective below is
to show that this can indeed happen, by constructing a model which
exhibits an extreme violation of the usual causality bounds --
allowing for exponentially accelerating signals of constant strength.
Other models compatible with the above intuition would naturally be
expected to show similar accelerations, albeit not necessarily
exponential ones. The model is constructed to violate the bounds in
the strongest possible fashion while still being solvable. It serves
as a proof of principle and as a worst case estimate for applications
of Lieb-Robinson bounds for the simulation of dynamics.

{\it Specific ``exchange interaction'' model. --}
We will pay special attention to the following model, defined for
bosons with spin 1 (so associated with the Hilbert space ${\cal H}={\cal
L}^2(\rr)\otimes \cc^3$).
We define for site $j$ the operators
$A_{j;k,l}  = |k,\uparrow\rangle \langle l,\downarrow|$
and $B_{j;k,l}  = |k,\downarrow\rangle \langle l,\downarrow|$.
The Hamiltonian is specified by
\begin{eqnarray*}
	f_{j,j+1}&=&
	\sum_{k,l=1}^\infty 
	(2k-1)
	\bigl(\mi
		A_{j;l,k}^\dagger B_{j+1; l,k} 
	+h.c.\bigr),\\
	g_j&=&2 \sum_{k=1}^\infty 
	(\mi k
	|k+1,\uparrow\rangle \langle k,\downarrow|
	+ h.c.)+ |0,\downarrow\rangle\langle 0,\downarrow|.
\end{eqnarray*}
Note that $f_{j,j+1}$ may be looked at as a variant of the familiar
exchange interaction. Clearly, $H$ is a legitimate 
bosonic Hamiltonian with translationally invariant nearest-neighbor
interactions. We will prove our claim three steps.

{\it 1. Mapping to an excitation Hamiltonian. --} 
To start with, $|0,\downarrow; \dots, 0,\downarrow
\rangle $ is an 
eigenstate of the Hamiltonian. If we now place a single 
excitation with spin $\uparrow$ 
at the first site---
so start with the initial state vector 
$|1,\uparrow;0,\downarrow;\dots;
0,\downarrow\rangle$---we see that time evolution will only
couple this to state vectors of the form 
\begin{equation*}
	||l\rangle\rangle = \left\{
	\begin{array}{ll}
	|0,\downarrow;
	\dots, 0,\downarrow;
	\frac{l+1}{2},\downarrow; 0,\downarrow; \dots, 0,\downarrow
	\rangle, &\text{ if $l$ odd,}\\
	|0,\downarrow;
	\dots, 0,\downarrow;
	\frac{l+2}{2},\uparrow; 0,\downarrow; \dots, 0,\downarrow
	\rangle,&\text{ if $l$ even,}
	\end{array} \right.
\end{equation*}
with particles 
at site $j=l/2+3/2$ and $j=l/2+1$, respectively
(for $l=0,\dots,2n-1$).
When considering only such excitations, 
we can hence pass to a new effective Hamiltonian 
$E$ with
Hilbert space ${\cal K}=\cc^{2n}$ and initial
condition $||0\rangle\rangle$,
\begin{equation*}
	E=\sum_{l=0}^{2n-1} \mi (l+1) 
	\left( ||l+1\rangle \rangle \langle\langle l||- ||l\rangle\rangle
	\langle \langle l+1|| \right).
\end{equation*}	
This Hamiltonian faithfully models the single 
excitation sector in the above sense. 
Note that $||l\rangle\rangle$ has now two roles:
It both refers to a position in 
the original lattice, as well as the particle number in
the original Hamiltonian. Also, to 
simplify notation later on,
we will at this point
pass to the half-open chain by setting $n=\infty$
\cite{CS}. 
The form of $E$ makes it manifest that the excitation experiences
stronger coupling coefficients as it moves along the lattice. This
alone is not sufficient to ensure an accelerated, directed movement:
convincingly, the ramping up in the coupling coefficient could cause
the excitation to scatter back. Maybe surprisingly, we will find
travelling solutions below.
  
{\it 2. Moments. --}
In this single excitation sector, time evolution corresponds to
$\rho(t) = e^{-\mi
t E}\rho(0)e^{\mi
t E}$, for states on
${\cal K}$.
It proves expedient to introduce the operators
\begin{eqnarray*}
	X=\sum_{l=0}^\infty (l+1)
	||l\rangle\rangle\langle \langle l|,\,
	P=\sum_{l=1}^\infty (l+1)
	\left( || l+1\rangle \rangle\langle\langle l|+ h.c. \right).
\end{eqnarray*}
Note that $X$ corresponds to a discrete position operator, measuring
twice the distance of the original model.
There is a lot of structure in this 
model: the commutation relations 
between these operators form a closed algebra. Indeed, 
\begin{eqnarray*}
	\mi
	[E,X]&=&P,\,\,\,\,\,
	\mi
	[E,P]= 4X - 2\id,
\end{eqnarray*}
the algebraic completion of which 
being that of the Lie-group $su(2)$.
Using the familiar Baker-Hausdorff formula, 
the Heisenberg picture
time evolution of $X$ under $E$ is given by
\begin{equation*}
	X(t)= e^{\mi tE}Xe^{-\mi tE}= 
	X+t[\mi E,X] + \frac{t^2}{2}[\mi E,[\mi 
	E,X]]+ \dots.
\end{equation*}
One obtains a closed-form expression for $X(t)$ by
exploiting the relations in the algebra ${\cal A}=\{E,X,P,\id\}$ to
iteratively solve these nested commutators. Explicitly
\begin{eqnarray*}
	X(t)&=& X+P\sum_{l=1}^\infty
	\frac{t^{2l-1}4^{l-1}}{(2l-1)!}
	+
	(4X-2\id)\sum_{l=1}^\infty \frac{t^{2l}4^{l-1}}{(2l)!}\nonumber\\
	&=& X-P \frac{1}{2}\sinh(2t)+
	\bigr(X-\frac{1}{2}\id\bigl) (\cosh(2t)-1),\nonumber
\end{eqnarray*}
For the time evolution of 
$X$ starting from the single excitation $||0\rangle\rangle$, 
we hence find 
\begin{equation*}
	\langle \langle 0|X(t) | 0\rangle\rangle =  
	\frac{1}{2}(1+\cosh(2t)).
\end{equation*}
Thus, the expectation value of $X$ is increasing exponentially in
$t$.  This fact alone, however, is not enough to show that we have
signaling: It could be that the excitation develops a long asymptotic
tail that leads to large first moments, but carries a small weight.
Thus, further information is needed. It turns out that knowledge of
the second moments $\langle \langle 0|| X(t)^2||0 \rangle\rangle -
\langle\langle 0 || X(t)|| 0 \rangle\rangle^2$ is sufficient to prove
signaling using the convex optimization ideas below.  An
analogous---if more tedious---calculation arrives at
\begin{eqnarray*}
	\langle\langle  0|| X(t)^2 ||0\rangle\rangle 
	&=&  \cosh^2(t) \cosh(2 t).	
\end{eqnarray*}

{\it 3. Hitting time from a convex optimization problem. --}
For some site $m>1$ define the {\it hitting operator}
\begin{equation*}
	T=\sum_{l=2m-1}^\infty
	||l\rangle\rangle\langle\langle l||.
\end{equation*}
So $P_1= \text{tr}[\rho(t)T]$ is the signal a distant observer 
$m$ sites away from the origin may receive, compared to $P_0=0$. 
Hence, the set $A=\{1\}$ is the single first site, 
whereas $B=\{m+1,\dots, \infty\}$ is the natural right part of the
chain. Let us set $M=2m-2$.
We will now bound this
expectation value by analytically solving a convex
optimization problem: 
\begin{eqnarray*}
	\text{minimize} && \sum_{l=M+1}^\infty p_l,\\
	\text{subject to} && \sum_{l=0}^\infty (l+1) p_l= a(t), 
	\,\,\,
	\sum_{l=0}^\infty (l+1)^2 X_l= b(t),
\end{eqnarray*}
$\sum_{l=0}^\infty p_l=1$, 
and $p_l\geq 0$ for all $l$, where $a(t)=(1+\cosh(2t))/2$
and $b(t)=\cosh^2(t) \cosh(2 t)$. 
So we minimize the signal, given first and second moments,
as a worst case analysis. This is an (infinite-dimensional)
 linear program, and hence an instance of a convex
 optimization problem. We can readily get a bound to the
 optimal solution by identifying a suitable solution to  the 
 Lagrange dual problem \cite{Convex}, which is
 \begin{eqnarray*}
	\text{maximize} && -d^T y,\\
	\text{subject to} && F^T y\geq -c,
\end{eqnarray*}
where $F_{1,j}= j$, $F_{2,j}=j^2$, $F_{3,j}= 1$
for  $j=1,2,\dots$, and $d(t)=( a(t),b(t),1)^T$. 
Also, $c=(0,\dots,0,1,1,\dots)^T$,
as a vector starting with $M$ zeros. The latter is an 
optimization problem over $y$. Any 
solution we can identify of the
dual will give a lower bound to the primal. It can be 
shown explicitly that the subsequent vector 
$y^*=(-2 (1+M)/M^3,
	1/M^4 ,(1+M)^2/M^2 -1)^T$
is always a feasible solution of the dual problem. 
Taking the time $t^\ast=\log(M)$ gives
\begin{eqnarray*}
	c^T x^\ast &\geq & -d^Ty^*=
	2 a(\log(M)) (1+M)/M^3 \nonumber\\
	&-&
	 b(\log(M))/M^4 -
	(1+M)^2/M^2+1=:g(M).
\end{eqnarray*}
For this function we have that
$\lim_{M\rightarrow\infty}
	g(M)=3/8$, and $g(M)>1/5$ for $M\geq 9$.
The solution of the dual will hence give rise to a lower
bound of our primal problem.
Hence, after a time $t^\ast$ logarithmic in $m$, a signal of constant
strength 
$\text{tr}[T\rho(t^\ast)]>1/5$ will have reached party $B$!
This means, of course---within the validity of the model---that 
we can in principle signal at any speed 
over arbitrary distances: The signal will not even decay,
and the Holevo-$\chi$ and the {\it classical information capacity}
of the associated quantum channel
are indeed constant. One can communicate with an
exponentially accelerating signal of
type (iii) in the above classification.

{\it Summary and Outlook. --} In this work, we have shown that in
systems of locally interacting bosons, excitations may accelerate and
carry information faster than any finite speed of sound.  Many results
on the simulatability of dynamics and ground-state properties of spin
systems (in particular using DMRG techniques) rely on
Lieb-Robinson bounds \cite{AreaReview}. Exceeding the usual
difficulties associated with reasoning about infinite-dimensional
systems, our examples imply further difficulties any generalization of
such results to bosonic systems has to deal with. This remains true
even if one restricts attention to low-energy sectors. Interestingly,
while analytical and numerical treatments of such models face
formidable difficulties, they may be simulated using analogue physical
systems, such as coupled cavity arrays \cite{Cavities} 
where massless polaritonic excitations are not subject to particle number
conservation.

{\it Acknowledgments. --}
This work has been supported by the EU (QAP, COMPAS, CORNER), 
the EPSRC, QIP-IRC, Microsoft Research, and the EURYI.

\section{Appendices}

For clarity of the argument, we present a few 
aspects of the above presentation in some more detail.
Note that this material is identical with the above one.

\subsection*{Appendix A: The spin Hamiltonian}

We consider a bosonic model with a spin-$1$ internal
degree of freedom. The local Hilbert space is hence
spanned by $\{{|k,\uparrow}\rangle, |{k,\downarrow}\rangle:k\in 
\nn_0\}$. We will be able to keep track of local
excitations in the following model. We take the Hamiltonian
\begin{eqnarray*}
	H&=& \sum_{j=1}^n
	\biggl(
	\sum_{k,l=1}^\infty (2k-1)
	\bigl(\mi
		|k,\downarrow\rangle \langle l,\uparrow|
		\otimes 
		| l,\downarrow\rangle \langle k,\downarrow|
	+h.c.\ \bigr) \nonumber\\
	&+& g_j\biggr),
\end{eqnarray*}
where the on-site interation is taken to be
\begin{equation*}
	g_j= 2 \sum_{k=1}^\infty 
	\bigl(\mi k 
	|k+1,\uparrow\rangle \langle k,\downarrow|
	+ h.c.
	\bigr)+ |0,\downarrow\rangle\langle 0,\downarrow|.
\end{equation*}
The hopping term is a variant of an exchange interaction
between neighboring sites (except from the spin, this would
exactly be a standard exchange interaction).
This Hamiltonian can also be written as
\begin{eqnarray*}
	H&=& \sum_{j=1}^n \biggl(
	\sum_{k,l=1}^\infty 
	(2k-1)
	\bigl(\mi
		A_{j;l,k}^\dagger B_{j+1; l,k} 
	+h.c.\bigr) +g_j\biggr),
\end{eqnarray*}
with the hopping operators at site $j$
being defined as
\begin{equation*}
	A_{j;k,l}  = |k,\uparrow\rangle \langle l,\downarrow|,\,\,\,
	B_{j;k,l}  = |k,\downarrow\rangle \langle l,\downarrow|,
\end{equation*}
The initial state vector is taken to be 
\begin{equation*}
	|\psi\rangle = |0,\downarrow\rangle^{\otimes n}.
\end{equation*}
This is obviously an eigenstate of the Hamiltonian. Now 
we excite site $1$, forming set $A=\{1\}$, by setting
it to $|1,\uparrow\rangle$, while keeping the rest of the chain
unaltered. We then see that the initial state vector
after the excitation
\begin{equation*}	
	|1,\uparrow\rangle|0,\downarrow\rangle^{\otimes (n-1)}
\end{equation*}
couples to
\begin{equation*}	
	|0,\downarrow\rangle
	|1,\downarrow\rangle 
	|0,\downarrow\rangle^{\otimes (n-2)},
\end{equation*}
and this to the former and to
\begin{equation*}	
	|0,\downarrow\rangle
	|2,\uparrow\rangle 
	|0,\downarrow\rangle^{\otimes (n-2)}.
\end{equation*}
Now the pattern is clear: All excitations are 
contained in 
\begin{eqnarray*}
	\text{span}
	&&
	\{ |0,\downarrow\rangle^{\otimes k}
	|k,\downarrow\rangle
	|0,\downarrow\rangle^{\otimes (n-k-1)},\\
	&&|0,\downarrow\rangle^{\otimes k}
	|k+1,\uparrow\rangle
	|0,\downarrow\rangle^{\otimes (n-k-1)}
	\},\nonumber
\end{eqnarray*}
and each basis vector---except at the ends of the chain---couples
to two further basis vectors. It is now a straightforward
exercise to verify that in this subspace of excitations, one
arrives at the above given effective Hamiltonian.

\subsection*{Appendix B: Convex optimization problem}

The matrix $F$ in the convex optimization problem 
defining our hitting time problem 
is given by
\begin{equation*}
F=\left[\begin{array}{ccccc}
	1 & 2 & 3 & 4 & \dots \\
	1 & 4 & 9 & 16 & \dots \\
	1 & 1 & 1 & 1 & \dots \\
\end{array}\right].
\end{equation*}
The first row captures the first moments, the second the second moments,
whereas normalization of the probability distribution
is enforced by the third one---together with the fact that all entries
of $x$ are positive.
The vector $c=(0,\dots, 0,1,\dots, 1)^T$, as a vector starting
with $M$ zeros. The primal problem can hence be written
as
\begin{eqnarray*}
	\text{minimize} & c^T x,\\
	\text{subject to} & F x = d,\\
	& x\geq 0,
\end{eqnarray*}
where $d(t)= (a(t), b(t), 1)^T$. 
This is a linear program in standard form, with a matrix
inequality constraint, and positive elements of the 
objective vector $x$. In this form it is easiest to identify
the dual problem. It is given by
\begin{eqnarray*}
	\text{maximize} & -d^T y,\\
	\text{subject to} & F^T y \geq -c,
\end{eqnarray*}
where now $y$ is not constrained to be positive.
The constraints on the vector
$y$ can equally be written as
\begin{equation}\label{Constraint}
	j y_1 + j^2 y_2 + y_3 + \delta_{j>M}\geq 0
\end{equation}
for all $j$, where $\delta_{j>M}$ takes the value $1$ if 
$j>M$ and is zero otherwise. The question is: Can we find
for each $M$ a choice of $(y_1,y_2,y_3)$ and a time $t$
such that we can meaningfully find a positive solution to the
dual problem?

We can take any solution to the dual problem. 
We choose
\begin{eqnarray*}
	y_1&=&-2 (1+M)/M^3,\\
	y_2&=&1/M^4 ,\\
	y_3&=&(1+M)^2/M^2 -1,
\end{eqnarray*}
and will verify that it is indeed a solution with the 
appropriate properties. The intuition behind this 
construction is as follows:
Consider the quadratic function
$f:[0,\infty)\rightarrow\infty$ as
\begin{equation*}
	f(x)= x y_1 +x^2 y_2 + y_3 .
\end{equation*}
This is identical with the left hand side of 
Eq.\ (\ref{Constraint}), up to being defined on $[0,\infty)$. 
We require that $f$ takes the value $0$ for the first time
exactly at $x=M$, and that $f(x)> -1$ for all $x$. 
Taking $y_2=1/M^4$ then gives rise to the above 
construction, satisfying 
Eq.\ (\ref{Constraint}) for all $M$ and all $j$. We have
$j y_1 + j^2 y_2 + y_3=0$ exactly at $j=M$, which is
a desirable feature to get a meaningful bound to the
original problem at hand. This is hence always a feasible 
solution to the dual problem. We will therefore obtain
a lower bound to the optimal solution of the primal 
problem $c^T x^\ast$ as
\begin{eqnarray*}
	c^T x^\ast &\geq &
	-d^T y\nonumber\\
	&=& 2 a(t) (1+M)/M^3 - b(t)/M^4 \nonumber\\
	&-&
	(1+M)^2/M^2+1.
\end{eqnarray*}
Choosing the time $t^\ast=\log(M)$, logarithmic in the distance,
we find
\begin{eqnarray*}
	c^T x^\ast &\geq &
	2 a(\log(M)) (1+M)/M^3 - b(\log(M))/M^4 \nonumber\\
	&-&
	(1+M)^2/M^2+1=:g(M).
\end{eqnarray*}
Now it is not difficult to see that indeed, in the limit of 
large $M$ we have
\begin{equation*}
	\lim_{M\rightarrow\infty}
	g(M)=3/8,
\end{equation*}
and that 
$g(M)>1/5$ for all $M\geq 9$. This means that for each
large $m$---and hence each large $M$---we can arrive at a constant signal for an appropriate constant signal.

\subsection*{Appendix C: Bose-Hubbard-type models}

In a sketchier fashion,
we will now briefly address the question whether even
Bose-Hubbard-type models
\begin{eqnarray*}
	H=\sum_{j=1}^n\left(
	\bigl( 
	a_j^\dagger a_{j+1}
	+ h.c.\bigr)+ h(a_j^\dagger a_j)\right),
\end{eqnarray*}
with some function $h:\rr^+\rightarrow \rr^+$ can
display such a behavior.
Indeed, it can be seen that if $h$ increases in $n$
sufficiently strongly, leading to highly interacting
particles, then there exists for each chain length $n$
an $N$ (polynomial in $n$) such that the 
initial state $|\psi\rangle=|N+1\rangle |N\rangle^{\otimes 
(n-1)}$---so an additional particle at site 
$A=\{1\}$ compared to $|N\rangle^{\otimes n}$---will 
lead to a signal at site $B=\{n\}$ that is at most polynomially
suppressed in $n$. The signal increases linearly in $N$.  Hence,
hopping models can also display violations of any finite bound on
propagation speeds, at least in the sense (i) of the above
classification.

For each chain length $n$, we will
consider a relevant $n$-dimensional
subspace of the original Hilbert space
\begin{equation*}
	{\cal L}_N = \text{span}
	\{|j_1,\dots, j_n\rangle|
	\exists k: j_k=N+1, j_l=N\, \forall l\neq k
	\},
\end{equation*}
for a suitable integer $N$, and denote with
${\cal L}^\perp_N={\cal H}\backslash {\cal L}_N$ the orthogonal
complement of the infinite-dimensional Hilbert space. 
The initial 
state vector of the chain of length $n$ is then taken to be
\begin{equation*}
	|\psi\rangle =|N+1\rangle |N\rangle^{\otimes (n-1)}
	\in {\cal L}_N.
\end{equation*}	 
This means that 
we fill up all number states with bosons at each site up to $N$, and
add an additional particle at site $A=\{1\}$ 
as a local excitation. 
We will now see that
we can then approximate the dynamics
of the chain within a spin chain 
restricted to a single excitation ${\cal L}_N$
arbitrarily well.
Define for site $j$ the operator
$A_j=|N\rangle\langle N+1|$, then the above
chain is approximated in its dynamics by the dynamics of
\begin{equation*}
	V=  (N+1)  
	\sum_{j=1}^n
	\left( A_j^\dagger A_{j+1}+ h.c.\right).
\end{equation*}
This dynamics is specifically simple, and we can see
that in a time $t^\ast$ that is linear in $N$
we will receive a 
signal at site $B=\{n\}$ that is at most 
polynomially
suppressed in $n$. This clearly
means that this signal will be received at an arbitrarily
velocity, if $N$ is sufficiently large.
To see that this mapping can be done to arbitrary approximation, 
we need the subsequent observation:

\begin{lemma}
For any Hermitian matrix $M$, partioned as
\begin{equation*}
	M= \left[
	\begin{array}{cc}
	A & B\\
	B^\dagger & C
	\end{array}
	\right].
\end{equation*}
we have that for all $t\in [0,\infty)$,
 \begin{equation*}
	\lim_{x\rightarrow\infty} \sup_{K\in {\cal M}_x}
	\| O \me^{\mi t K}O^\dagger - \me^{\mi tA} \|
	=0
\end{equation*}
where 
\begin{equation*}
	{\cal M}_x=\left \{ M+\left[
	\begin{array}{cc}
	0 & 0\\
	0 & P
	\end{array}
	\right]: P\geq x \id \right\},
	\,\, O= \left[
	\begin{array}{cc}
	\id & 0\\
	\end{array}
	\right].
\end{equation*}
\end{lemma}

Here, $\|.\|$ denotes the operator matrix or
vector norm. 
To prove this, note that 
\begin{equation*}
	\sum_{j=1}^d
	\lambda^\uparrow_j
	\left(
	\left[
	\begin{array}{cc}
	A & B\\
	B^\dagger & C+P
	\end{array}
	\right]
	\right)=\inf_Q \text{tr}
	\left[
	Q \left[
	\begin{array}{cc}
	A & B\\
	B^\dagger & C+P
	\end{array}
	\right]
	\right],
\end{equation*}
where the infimum is taken over
$Q$ that are projectors of rank $d$, when 
$A$ is a $d\times d$-matrix. This infimum
exists for any $M\in {\cal M}_x$, call it $Q_M$. Since
$\|A\|=c$ is constant, we have that
\begin{equation*}
	\lim_{x\rightarrow\infty}
	\sup_{M\in {\cal M}_x}
	\left \|Q_M-\left[
	\begin{array}{cc}
	\id & 0\\
	0 & 0 \\
	\end{array}
	\right]
	\right\|=0.
\end{equation*}
Hence, since $|\me^{\mi t \lambda_j^\uparrow(M)}|=1$,
the assertion follows.

Subsequently, 
the submatrix $A$ will be identified with 
$H$ restricted to ${\cal L}_N$, whereas 
$C$ is the restriction to the orthogonal
complement ${\cal L}_N^\perp$. Using this
observation, we see that for each 
$\varepsilon>0$ we can find a 
function $h:\rr^+\rightarrow \rr^+$ such that 
the following property is satisfied: 
For each chain length $n$ there exists an $N$
such that for the above hitting time $t^\ast$
\begin{equation*}
	\| \me^{-\mi t^*  H}|\psi\rangle - \me^{-\mi t^* V}|\psi\rangle\|<
	\varepsilon
\end{equation*}
for all $t\in [0,t^\ast]$. But this means that we can
signal with this chain, in the sense of (iii) in the above
classification, as the class of
restricted models has this property.


\begin{thebibliography}{99}
\bibitem{HarmonicFootnote}
	Harmonic versions of Lieb-Robinson
	bounds have recently been derived \cite{Schlein}, and 
	such chains have also been explored for quantum state 
	transfer \cite{Harmonic}.
\bibitem{ChebRemark}
	By Chebychev's inequality, the probability of 
	observing an energy
	value exceeding $t E_0$ when performing a measurement on
	$|\psi\rangle$ is smaller than $1/t^2$.
\bibitem{CS}
	With greater notational effort, the statements involving infinite
	chains can be made more precise, by e.g., passing to the
	Heisenberg picture and employing a $C^*$-algebraic language.
\end{thebibliography}
\end{document}